\documentclass[12pt]{iopart}

\usepackage{epsfig}
\usepackage{psfrag}
\usepackage{graphicx}
\usepackage{dcolumn}
\usepackage{bm}
\begin{document}

\title{Rapidity losses in heavy-ion collisions from AGS to RHIC energies}
\author{F. C. Zhou$^{1, 2}$
, Z. B. Yin$^{1, 2}$ and D. C. Zhou$^{1, 2}$
}

\address{$1$ Institute of Particle Physics, Huazhong Normal
University, Wuhan 430079, China}
\address{$2$  Key Laboratory of Quark $\&$ Lepton Physics (Huzhong
Normal University), Ministry of Education, China}

\ead{zbyin@mail.ccnu.edu.cn}

\begin{abstract}

We study the rapidity losses in central heavy-ion collisions from AGS 
to RHIC energies with the mean rapidity determined from the 
projectile net-baryon distribution after collisions. 
The projectile net-baryon distribution in the full rapidity range
was obtained by removing the target 
contribution phenomenologically at forward rapidity region from the 
experimental net-baryon measurements and taking into account the projectile 
contribution at backward rapidity region.
Based on the full projectile net-baryon distributions, calculation results 
show that the rapidity loss stops increasing from the SPS top energy 
to RHIC energies, indicating that baryon transport does not 
depend strongly on energy at high energies. 

\end{abstract}


\section{Introduction}
\label{sec:intr}

The ultimate goal of studying (ultra-)relativistic heavy-ion collisions is
to search for the signatures of possible quark-gluon plasma (QGP) 
formation and 
explore the properties of QGP and the phase transition from hadronic 
matter to de-confined quark matter~\cite{BRAH05,PHOB05,STAR05,PHEN05}. 
In heavy-ion collisions, 
the colliding nuclei lose part of their kinetic energy for 
the possible formation of QGP and new particle 
production. The degree of kinetic energy loss 
is customarily quantified by the average rapidity loss, 
\begin{equation}
<\delta y> = y_p - <y_b> ,
\label{eq:deltay}
\end{equation}
where $y_p$ is the rapidity of the projectile before the collisions 
and $<y_b>$ is the mean rapidity of projectile net-baryons 
after the collisions~\cite{Busz84,Vide95}. However, experimentally it is not 
possible to distinguish between baryons originating
from the target or from the projectile. Thus, phenomenological 
models have to be used in order to extract the pure projectile net-baryon 
rapidity distribution from experimental data. 

Rapidity losses in heavy-ion collisions were measured at different 
energies from AGS, SPS to RHIC~\cite{AGS01,SPS99,BRAH04,BRAH09} 
with aims of determining the degree of nuclear stopping and energy 
density built up in the reaction zone. At AGS energies, it was 
demonstrated by the E917 collaboration that 
the net-baryon distributions after heavy-ion collisions 
could be well described with double gaussians~\cite{AGS01}. The mean rapidity 
losses were then correctly estimated by using Eq.~(\ref{eq:deltay}) with 
the $<y_b>$ determined from the gaussian centered at positive rapidity, 
which was assumed to correspond to the projectile baryon distribution.      
At SPS and RHIC energies the mean rapidity losses were calculated 
without distinguishing the origin of the net-baryons but the 
mean rapidity value from the mid-rapidity to $y_p$ was taken 
as the $<y_b>$~\cite{SPS99,BRAH04,BRAH09}. 
This makes the comparison of rapidity 
losses at different energies complicated when the contribution 
of target baryon to the net-baryon distribution at the rapidity region 
above mid-rapidity showed a strong energy dependence~\cite{BRAH09}.
In order to study the energy dependence of the rapidity loss, it is thus
necessary to examine the sensitivity of rapidity loss 
value to the target baryon contribution.

In this paper, we first extract in section~\ref{sec:proj} 
the pure projectile net-baryon distribution from experimental 
measurements of net-baryon distributions based on phenomenological model 
descriptions of target baryon contribution to the net-baryon yields at 
positive rapidity in the center-of-mass system. Then we 
calculate the average rapidity loss in the most 
central heavy-ion collisions at different energies with the obtained pure 
projectile net-baryon distribution and show its energy dependence 
in section~\ref{sec:loss}. Finally, we conclude with 
a short summary in section~\ref{sec:con}.  

\section{The projectile net-baryon rapidity distribution}
\label{sec:proj}

The projectile baryons peak at $y_p$ before the heavy-ion collision, 
while after the collision the projectile baryon distribution can extend 
from target rapidity to the projectile rapidity~\cite{Vide95}. Thus, 
to obtain the average projectile baryon rapidity $<y_b>$ after the collision, 
the integration should be carried out from the target rapidity to 
the projectile rapidity. For symmetrical heavy-ion collisions, 
we express the $<y_b>$ in the center-of-mass system as
\begin{equation}
<y_{b}>=\frac{2}{N_{\rm{part}}}\int_{-y_{p}}^{y_{p}}{y\frac{dN^{B-\bar{B}}}{dy}dy},
\label{eq:y_b}
\end{equation}

\noindent
where $\frac{dN^{B-\bar{B}}}{dy}$ is the projectile net-baryon 
rapidity distribution and $N_{\rm{part}}$ is the number of participating 
baryons in the collisions. In collider experiments one can choose any of 
the two incoming beams as the projectile. In this paper, we consider
the nuclear beam in positive rapidity direction as the projectile. 

As it is experimentally impossible to distinguish baryons originating 
from the target or from the projectile, one has to rely on 
phenomenological models to extract the projectile net-baryon distribution.
We use the same approach as described in Ref.~\cite{BRAH09}, where 
the target contribution at positive rapidity region is 
estimated as the average of two different rapidity dependences : 
(1) a simple exponential form $\exp(-y)$~\cite{Busz84}
and (2) a gluon junction motivated form 
$\exp (-\frac{y}{2})$~\cite{Kope89}. For a 
symmetrical colliding system, the projectile baryons should
comprise half of the whole net-baryons at mid-rapidity, the subtraction of 
the target contribution based on the phenomenological functions is therefore 
constrained by the net-baryon measurements at mid-rapidity.
At AGS energies, as there is a non-negligible target contribution 
around $y_{p}$, the double gaussians form used in Ref.~\cite{AGS01} 
is found to be one of the best forms to describe 
the experimental net-baryon data and thus
we will adopt their results on rapidity losses when we discuss on the 
energy dependence in the following section. 

\begin{figure}
\centering
\includegraphics[width=10cm]{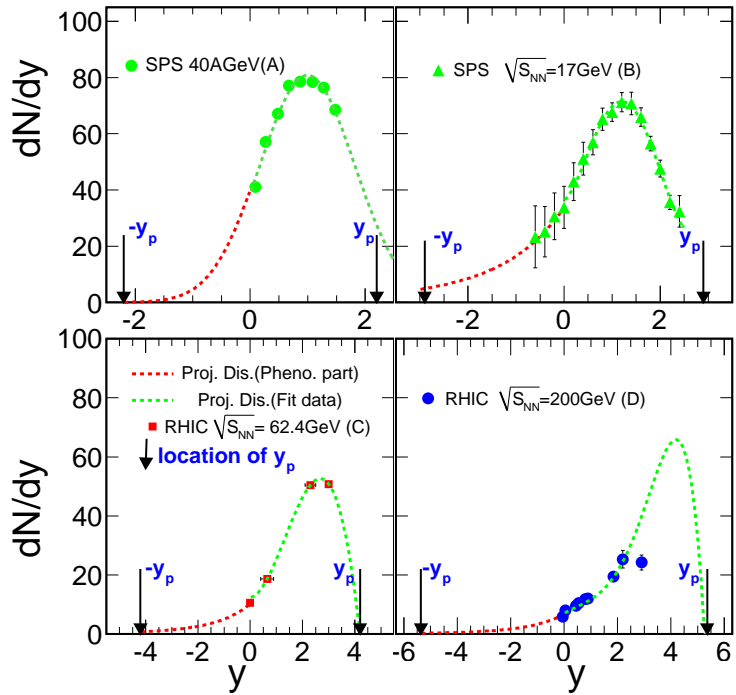}
\caption{The projectile net-baryon rapidity distribution 
in the center-of-mass system. The red dashed lines indicate 
the projectile contribution at target rapidity region. 
The data points show the resulted projectile 
baryon density after subtracting phenomenologically 
the target contribution from the experimental measured net-baryon 
data at positive rapidity region. The green dashed lines are curves 
used to fit the data points. The arrows indicate the values of 
the beam rapidities. Panel (A), (B), (C) and (D) correspond 
respectively to the most 
central heavy-ion (Pb+Pb or Au+Au) collisions 
at SPS with beam energy of 40 AGeV~\cite{Stro09}, 
SPS at $\sqrt{s_{NN}} = 17$ GeV~\cite{SPS99}, 
RHIC at $\sqrt{s_{NN}} = 62.4$ GeV~\cite{BRAH09}, 
and RHIC at $\sqrt{s_{NN}}  = 200$ GeV~\cite{BRAH04}. }
\label{fig:netb}
\end{figure}

In Fig.~\ref{fig:netb} the data points represent the resulted projectile
baryon densities after phenomenologically subtracting 
the target contribution from the experimental measured net-baryon
data at positive rapidity region. The green dashed lines are curves used 
to fit the data points considering the conservation of baryon number 
as discussed below. The arrows indicate the values of beam rapidities.
Panel (A), (B), (C) and (D) correspond to the most
central heavy-ion (Pb+Pb or Au+Au) collisions at SPS with beam 
energy of 40 AGeV~\cite{Stro09}, SPS at $\sqrt{s_{NN}} = 17$ GeV~\cite{SPS99},
RHIC at $\sqrt{s_{NN}} = 62.4$ GeV~\cite{BRAH09},
and RHIC at $\sqrt{s_{NN}}  = 200$ GeV~\cite{BRAH04}, respectively. 
Due to the projectile-target symmetry, the projectile baryon 
distribution at negative rapidity region should be the same as the 
target contribution at the positive rapidity region. These two together
should form an even function in the center-of-mass system. 
In the figure the projectile contribution at negative rapidity region 
is indicated with red dashed lines. Thus, merging the red and 
green dashed lines at mid-rapidity 
gives the net-baryon distribution from projectile only 
in the whole rapidity range from $-y_p$ to $y_p$. 
It is worthwhile to mention that, 
in order to conserve the baryon number, the fit 
to the data points was done in such a manner that the total baryon number 
obtained by integrating the projectile net-baryon 
distribution from $-y_p$ to $y_p$ 
amounts to half of the number of participating baryons.

\begin{figure}
\centering
  \includegraphics[height=9cm,width=10cm]{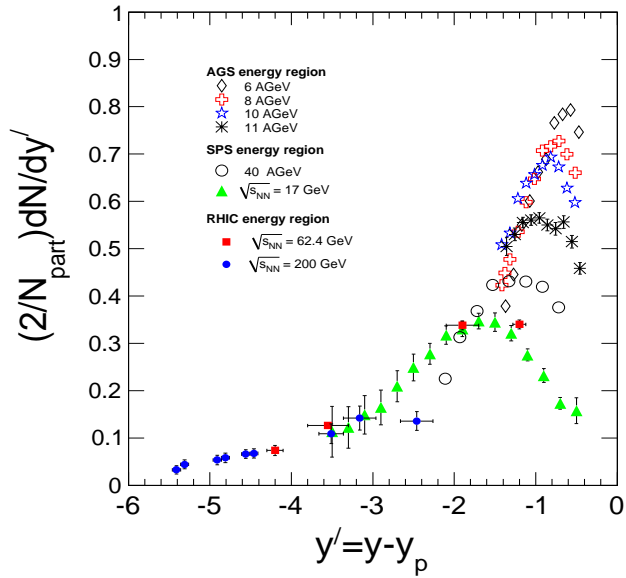}
  \caption{ Normalized projectile net-baryon rapidity density 
$\frac{2}{N_{part}}\frac{dN^{B-\bar{B}}}{dy^{\prime}}$ from AGS, SPS 
to RHIC energies in the frame of projectile.}
\label{fig:norm}
\end{figure}

To compare the projectile net-baryon distributions 
at different energies, Fig.~\ref{fig:norm} depicts a compilation of 
the projectile net-baryon rapidity density normalized to 
the number of participants, 
$\frac{2}{N_{part}}\frac{dN^{B-\bar{B}}}{dy^{\prime}}$, 
from AGS, SPS to RHIC 
energies in the rapidity frame of the projectile, $y^{\prime} = y - y_p$.
It is evident that the rapidity distribution peaks at 
lower $y^{\prime}$ values with higher colliding 
energies from AGS to top SPS energy. This indicates the mean rapidity loss 
increases with beam energy. 
However, the peak position at RHIC energy of $\sqrt{s_{NN}} = 62.4$ GeV 
does not show a significant difference from that at $\sqrt{s_{NN}} = 17$ GeV, 
implying a possible different mechanism of baryon transport and 
nuclear reaction~\cite{Bass99,Soff03,Bass03} above the top SPS energy 
compared to lower energies.

\section{The energy dependence of rapidity loss}
\label{sec:loss}

\begin{figure}
\centering
  \includegraphics[height=10cm,width=10cm]{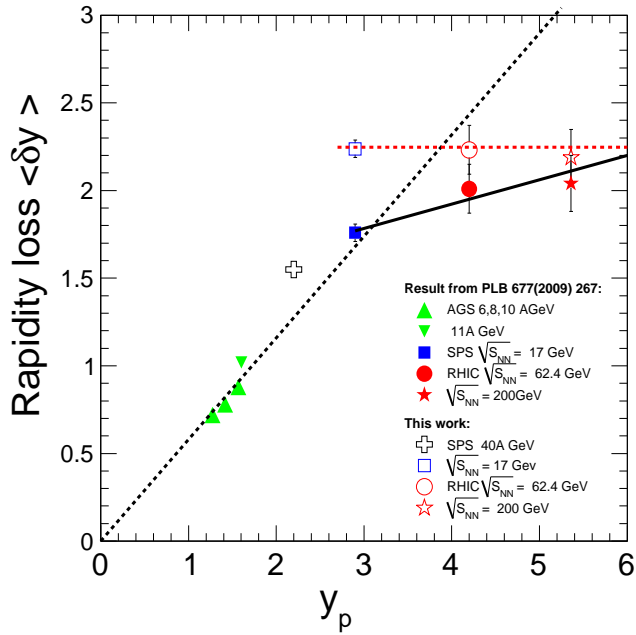}
  \caption{Rapidity losses in the most central heavy-ion (Au, Pb) 
collisions from AGS, SPS to RHIC as a function of beam rapidity. 
The black dashed line is a linear fit to the AGS energy rapidity 
loss as discussed in Ref.~\cite{Vide95}, while the red dashed 
horizontal line is used to guide your eyes.
The filled symbols are results published in ref.~\cite{BRAH09} and 
open symbols are results from this work. }
\label{fig:raploss}
\end{figure}

With the projectile net-baryon distribution as shown in Fig.~\ref{fig:netb},
we can calculate the rapidity losses at SPS and RHIC energies by using 
Eq.~(\ref{eq:deltay}) and Eq.~(\ref{eq:y_b}). The results are shown 
as open symbols in Fig.~\ref{fig:raploss} and compared to the published 
ones in Ref.~\cite{BRAH09}, in which the target 
contribution was not subtracted and the $<y_b>$ was evaluated from 
mid-rapidity to beam rapidity. It is evident that the correction on 
the rapidity loss value is largest at SPS energy and decreases with  
energy because both the target contribution at positive rapidity 
and the projectile contribution at negative rapidity to the net-baryon 
distribution decreases with increase of energy. As a comparison 
to previous results, another significant feature of our results 
is that the rapidity loss 
does not increase with energy above $\sqrt{s_{NN}} = 17$ GeV. 
This would indicate that the reported results in Ref.~\cite{BRAH09}
on the beam rapidity dependence of the rapidity loss can be 
misleading because the true rapidity loss value depends strongly on 
the target contribution to the net-baryon distribution 
at positive rapidity region.
As the target contribution decreases with increase of energy, 
one would expect the target contribution to become negligible at LHC energy.
Thus, it is very interesting to measure the rapidity loss value at LHC 
to examine the systematic trends of rapidity losses at high energies.   

\begin{figure}
\centering
  \includegraphics[height=10cm,width=10cm]{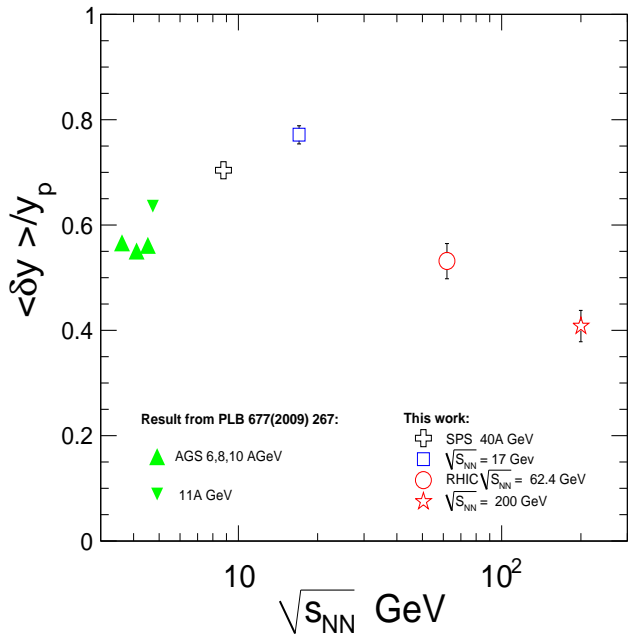}
  \caption{The relative mean rapidity loss as a function of the 
colliding energy in the center-of-mass system.}
\label{fig:rel_yloss}
\end{figure}

To explore the energy dependence of the nuclear stopping power, 
Fig.~\ref{fig:rel_yloss} depicts the relative rapidity loss 
as a function of the 
collision energy in the center-of-mass system. 
The relative rapidity loss shows an increase with energy from AGS to SPS 
but then a decrease above SPS energy, indicating that the largest
stopping power is achievable around the top SPS energy.   

\section{Conclusions}
\label{sec:con}

In this paper, we investigate a possible correction to the published results of
rapidity loss by extracting the projectile net-baryon distribution 
in the whole rapidity range from the experimental measurements of 
net-baryon distribution with help of phenomenological models on the target
contribution to the net-baryon density at positive rapidity.
This energy dependent correction leads to a different 
rapidity dependence of the mean rapidity loss compared 
to the results reported in Ref.~\cite{BRAH09}.
The mean rapidity loss starts to saturate at energy 
of $\sqrt{s_{NN}} = 17$ GeV and the largest stopping power is reached 
at the top SPS energy. This indicates that sufficiently high energy collisions
will get ``transparent'' as assumed by Bjorken~\cite{Bjor83} 
and will result in approximately net-baryon-free region at mid-rapidity. 
As the target contribution to the net-baryon density 
at positive rapidity region is expected to be negligible at LHC energies, 
we propose to measure the rapidity loss at LHC energies so as to verify or 
falsify the systematic trends of the beam rapidity dependence 
of the rapidity loss as observed in this paper. This measurement 
can also help to distinguish between different proposed phenomenological 
mechanisms of initial coherent multiple interactions and baryon transport 
in heavy-ion collisions at various beam energies~\cite{Busz84,Kope89}. 
 
\section*{Acknowledgments}
This work is supported by National Natural Science Foundation of
China with Grant Nos. 10975061, 10875051 and 10635020, 
by the National Key Basic Research Program of China 
under Grant No. 2008CB317106, and by the Chinese Ministry of 
Education under Key Grant Nos. 306022 and IRT0624.

\end{document}